\documentclass[aps,pra,twocolumn,preprintnumbers]{revtex4}
\usepackage{amsmath,amssymb,bm,graphicx,subfigure}

\begin{document}

\title{Effective low-dimensional Hamiltonian for strongly interacting
atoms in a transverse trap}
\author{J. P. Kestner, L.-M. Duan}
\affiliation{FOCUS center and MCTP, Department of Physics, University of Michigan, Ann
Arbor, MI 48109}

\begin{abstract}
We derive an effective low-dimensional Hamiltonian for strongly
interacting ultracold atoms in a transverse trapping potential near
a wide Feshbach resonance. The Hamiltonian includes crucial
information about transverse excitations in an effective model with
renormalized interaction between atoms and composite dressed
molecules. We fix all the parameters in the Hamiltonian for both
one- and two-dimensional cases.
\end{abstract}

\maketitle
\section{Introduction}
Cold atomic gases strongly confined via optical lattice techniques
along one or two dimensions and placed in an external magnetic field
tuned near a Feshbach resonance provide the exciting possibility to
study low-dimensional (low-D) strongly correlated physics in a
controllable fashion \cite{1,2}. For weakly interacting atoms in a
transverse trap, it is easy to get a low-dimensional Hamiltonian by
projecting the atomic wave function to the lowest trap mode.
However, the situation gets much more complicated for strongly
interacting atoms. It is known that virtual excitation of the
transverse trap modes during the atomic collisions could lead to
interesting renormalization of the atomic scattering lengths in low
dimensions \cite{3,4,5}. Furthermore, independent of the detuning
from the Feshbach resonance, the atoms can always form dimers
(two-body bound states) as their ground state configuration in a
transverse trap \cite{3,4,5,6,7,Kestner}. Because of the formation
of dimers, the problem does not simply reduce to atomic scattering
with a renormalized scattering length, but instead, we have to take
into account the \textit{real population} of the atoms in transverse
excited levels, which remains significant even for an extremely
strong transverse trap \cite{Kestner}.

In this paper, we construct the low-D (both 1D and 2D) effective
Hamiltonian by taking into account the atomic population in the
transverse excited modes. To describe the transverse excitations,
one can introduce a dressed molecule \cite{8}, which is part of the
dimer state excluding the atomic population in the lowest transverse
level. We observe that for realistic atomic densities, the structure
of the dressed molecule is largely fixed by the two-body physics and
thus almost density independent. This allows us to derive the
renormalized interaction between the atoms and the dressed
molecules, which gives the effective many-body Hamiltonian. This
effective Hamiltonian reproduces the correct two-body state, and
from that, we also fix all the parameters in the Hamiltonian at any
detuning of the magnetic field.
\section{Exact Hamiltonian}
We consider an atomic gas free in $D$ dimensions ($D=1,2$) and
trapped by a ($3-D$)-dimensional harmonic potential of frequency
$\omega $ along the transverse direction. (A single well of the
lattice potential can be well approximated with a harmonic trap.) We
use the conventional two-channel field theory for Feshbach resonance
\cite{9}, although one could also use a single-channel approach with
an energy-dependent pseudopotential \cite{10}.  For atoms of mass
$m$ possessing internal states $\sigma =\{\uparrow ,\downarrow \}$,
with contact interactions, the Hamiltonian is
\begin{multline}\label{eq:Hr}
H = \sum_{\sigma = \uparrow, \downarrow} \int d^{3}\mathbf{r}
\Psi_{\sigma}^{\dagger} \left( - \frac{1}{2}\nabla^2 + \frac{1}{2}
\sum_{i=1}^{3-D} x_i^2 \right) \Psi_{\sigma}
\\
+ \int d^{3}\mathbf{r} \Phi^{\dagger} \left( - \frac{1}{4}\nabla^2 +
\sum_{i=1}^{3-D} x_i^2 + \nu_b \right) \Phi
\\
+ g_b \int d^{3}\mathbf{r} \left( \Psi_{\uparrow}^{\dagger}
\Psi_{\downarrow }^{\dagger} \Phi + h.c. \right) + U_b \int
d^{3}\mathbf{r} \Psi_{\uparrow }^{\dagger} \Psi_{\downarrow
}^{\dagger} \Psi_{\downarrow} \Psi_{\uparrow}
\end{multline}
where $\Psi \left( \mathbf{r}\right) $ is the atomic field operator
and $\Phi \left( \mathbf{r}\right) $ is the molecular field
operator. Above, and throughout this paper, we use dimensionless
quantities with all energies in units of $\hbar \omega$ and all
lengths in units of the trap length scale, $a_{t}=\sqrt{\hbar
/m\omega }$. The dimensionless parameters are defined as follows:
$\nu _{b}$ is the bare detuning (in units of $\hbar \omega $),
$g_{b}$ is the bare atom-molecule coupling rate (in units of $\hbar
\omega a_{t}^{3/2}$), and $U_{b}$ is the bare background atomic
scattering rate (in units of $\hbar \omega a_{t}^{3}$). The bare
parameters are related to the physical ones via the standard
renormalization relations:
\begin{gather}
U_{c}^{-1}=\int \frac{d^{3}\mathbf{k}}{\left( 2\pi \right)^3}
\frac{1}{ 2E_{\mathbf{k}}}\,,\quad \Gamma ^{-1}=1-U_{p}\,U_{c}^{-1},
\notag
 \\
U_{b}=\Gamma U_{p}\,,\quad g_{b}=\Gamma g_{p}\,,\quad \nu _{b}=\nu
_{p}+\Gamma g_{p}^{2}U_{c}^{-1} \,, \label{eq:renormalization}
\end{gather}%
where the subscript $p$ denotes physical parameters, $E_{\mathbf{k}}
\equiv \mathbf{k}^2/2$, and the integral is taken in three
dimensions with an explicit energy cutoff $E_{c}$ (which finally
goes to infinity) imposed on two dimensions, so
$U_{c}^{-1}=\sqrt{E_{c}}/2^{3/2}\pi $. The physical parameters
$g_p$, $U_p$, $\nu_p$ are determined from the scattering data as
$U_p = 4\pi a_{bg}/a_t$, $g_p = \sqrt{4\pi \mu_{co}W|a_{bg}|/a_t
\hbar \omega}$, and $\nu_p = \mu_{co} \left( B - B_0 \right)/ \hbar
\omega$ ($\mu_{co}$ is the difference in magnetic moments between
the two channels), where we have assumed that the s-wave scattering
length near resonance has the form $a_s = a_{bg} \left( 1 - \frac{W}
{B-B_0} \right) $, with $a_{bg}$ as the background scattering
length, $W$ as the resonance width, and $B_{0}$ as the resonance
point.

Expanding the field operators $\Psi \left( \mathbf{r}\right) $ and
$\Phi \left( \mathbf{r}\right) $ in terms of trap eigenmodes in the
trapped dimensions and plane waves in the untrapped dimensions
yields
\begin{multline}
H=\sum_{\mathbf{mk}\sigma }\left( \epsilon _{\mathbf{k}}+\varepsilon _{%
\mathbf{m}}\right) a_{\mathbf{mk}\sigma }^{\dagger }a_{\mathbf{mk}\sigma }
\label{eq:H} \\
+\sum_{\mathbf{mk}}\left( \epsilon _{\mathbf{k}}/2+\varepsilon _{\mathbf{m}%
}+\nu _{b}\right) b_{\mathbf{mk}}^{\dagger }b_{\mathbf{mk}} \\
+\frac{g_b}{L^{D/2}}\sum_{\substack{ \mathbf{mnp}
\\ \mathbf{kq}}}\gamma _{%
\mathbf{mnp}}\left( a_{\mathbf{m,k+q,}\uparrow }^{\dagger }a_{\mathbf{n,-k,}%
\downarrow }^{\dagger }b_{\mathbf{pq}}+\text{h.c.}\right)  \\
+\frac{U_b}{L^D}\sum_{\substack{ \mathbf{mn%
} \mathbf{m^{\prime }n^{\prime }} \\ \mathbf{kk^{\prime }q}}}%
\gamma _{\mathbf{mn}}^{\mathbf{m^{\prime }n^{\prime }}}a_{\mathbf{%
m,k+q,}\uparrow }^{\dagger }a_{\mathbf{n,-k,}\downarrow }^{\dagger
}a_{\mathbf{n^{\prime },-k^{\prime },}\downarrow
}a_{\mathbf{m^{\prime },k^{\prime }+q,}\uparrow }
\end{multline}%
where $\mathbf{m}$ indexes trap eigenmodes $\{m_{i}\}$, $i=1,\ldots
,3-D$, and $\mathbf{k}$ denotes the wave vector in the untrapped
dimensions $\{k_{j}\}$, $j=1,\ldots ,D$. The operators
$a_{\mathbf{mk}\sigma }$ and $b_{\mathbf{mk}}$ represent the
corresponding atomic and Feshbach molecular modes, respectively. The
plane wave energy $\epsilon _{\mathbf{k}}$ and the mode
energy $\varepsilon _{\mathbf{m}}$ are given by $\epsilon _{\mathbf{k}%
}= \sum_{j=1}^{D}k_{j}^{2}/2$, $\varepsilon _{\mathbf{%
m}}=\sum_{i=1}^{3-D}m_{i}$ [for convenience we neglect the constant
energy $\left(3-D\right)/2$ in $\varepsilon _{\mathbf{m}}$ as we
will measure the two-body bound state energy with respect to the
continuum threshold $3-D$]. $L$ is the dimensionless quantization
length in the untrapped dimensions. The form factors appearing in
Eq. \eqref{eq:H} are given by
\begin{equation}\label{eq:gamma}
\gamma _{\mathbf{mnp}} = 2^{\left( 3-D\right)
/4}\int d\mathbf{r}\langle \mathbf{m}|\mathbf{r}\rangle \langle \mathbf{n}|%
\mathbf{r}\rangle \langle \sqrt{2}\mathbf{r}|\mathbf{p}\rangle ,
\end{equation}%
\begin{eqnarray}
\gamma _{\mathbf{mn}}^{\mathbf{m^{\prime }n^{\prime }}} &=&\int d%
\mathbf{r}\langle \mathbf{m}|\mathbf{r}\rangle \langle \mathbf{n}|\mathbf{r}%
\rangle \langle \mathbf{r}|\mathbf{m^{\prime }}\rangle \langle \mathbf{r}|%
\mathbf{n^{\prime }}\rangle   \notag \\
&=&\sum_{\mathbf{p}}\gamma _{\mathbf{mnp}}\gamma _{\mathbf{m^{\prime
}n^{\prime }p}}^{\ast },
\end{eqnarray}%
with
\begin{equation}\label{eq:phi}
\langle \mathbf{r}|\mathbf{m}\rangle =\prod_{i=1}^{3-D}\frac{%
e^{-r_{i}^{2}/2}}{\pi ^{1/4}\sqrt{2^{m_{i}}m_{i}!}}%
\;H_{m_{i}}\!\left( r_{i}\right) ,
\end{equation}%
where $H_{n}\!\left( x\right) $ is the Hermite polynomial.
\section{Effective Hamiltonian}
The above Hamiltonian is extremely complicated and hard to solve
directly. We expect it should reduce to some low-dimensional
effective Hamiltonian when the trap confinement is strong enough. In
particular, we assume that the many-body energy scale, characterized
by the density dependent part of the chemical potential $\mu _{\rho
}$, is much less than the trap energy $\hbar \omega $. This is
equivalent to assuming that the 1D (or 2D) atomic density $\rho \ll
1$ measured in units of $a_{t}^{-1}$\ (or $ a_{t}^{-2}$), which is
typically the case for realistic systems. Under this condition, when
the atoms are far apart, they should stay in the transverse ground
level to minimize the energy. The transverse excited levels get
populated only when the atoms come close to strongly interact with
each other. However, since $\rho \ll 1$ and there is no $n$-particle
bound state with $n\geq 3$ for two-component fermions, it is rare
for three or more atoms to come close (two atoms can approach each
other as there exists a bound dimer state at any detuning $\nu _{p}$
from the Feshbach resonance \cite{Kestner}). From this argument, we
see that the atomic distribution in the transverse excited modes is
determined by the short-range physics, where the latter is fixed
through the two-body solution. The dimer state of the Hamiltonian
$H$ with binding energy (relative to the continuum threshold, $3-D$)
$-E_{2B} > 0$ and momentum $\mathbf{q}$ can be written in the form
$\left\vert \Psi _{2}\right\rangle =\Psi _{2\mathbf{q}}^{\dagger
}\left\vert 0\right\rangle $, where $\left\vert 0\right\rangle $
denotes the vacuum and $\Psi _{2\mathbf{q}}^{\dagger }=\beta
b_{\mathbf{0q}}^{\dagger
}+\sum_{\mathbf{mnk}}\frac{\eta _{\mathbf{mn}}}{E_{2B}-2\epsilon _{\mathbf{k}%
}-\epsilon _{\mathbf{q}}/2-\varepsilon _{\mathbf{m}}-\varepsilon _{\mathbf{n}%
}}a_{\mathbf{m,k+q/2,}\uparrow }^{\dagger
}a_{\mathbf{n,-k+q/2,}\downarrow }^{\dagger }$ with the coefficients
$\beta $ and $\eta _{\mathbf{mn}}$ given in Ref. \cite{Kestner},
although the actual expressions are not important for the purposes
of this discussion. We thus construct the dressed molecular modes
$d_{\mathbf{q}}^{\dagger }$ with the same
expression as $\Psi _{2\mathbf{q}}^{\dagger }$, but excluding the $%
\mathbf{m=n=0}$ term in the summation (correspondingly normalized).
These modes $d_{\mathbf{q}}^{\dagger }$ capture the short-range
physics and their structure should be basically independent of the
atomic density. The coupling between $d_{\mathbf{q}}$ and the atomic
modes $a_{\mathbf{0k}\sigma }$ (with a simplified notation as
$a_{\mathbf{k}\sigma }$) in the open channel can be approximated
with a contact interaction since $d_{\mathbf{q}}$ is tightly
confined in space to a volume on the order of $a_t^3$ in the deep
BCS regime and even smaller in the crossover and BEC regimes
\cite{Kestner}. The general effective Hamiltonian for
$d_{\mathbf{q}}$ and $a_{\mathbf{k}\sigma }$ then takes the form
\begin{multline}
H_{\text{eff}}=\sum_{\mathbf{k}\sigma }\epsilon
_{\mathbf{k}}a_{\mathbf{k} \sigma }^{\dagger }a_{\mathbf{k}\sigma
}+\sum_{\mathbf{q}}\left( \epsilon _{ \mathbf{q}}/2+\lambda
_{b}\right) d_{\mathbf{q}}^{\dagger }d_{\mathbf{q}}
\label{eq:H_eff} \\
+\frac{\alpha _{b}}{L^{D/2}} \sum_{\mathbf{kq}} \left(
a_{\mathbf{k+\frac{q}{2 },}\uparrow }^{\dagger
}a_{\mathbf{-k+\frac{q}{2},}\uparrow }^{\dagger }d_{
\mathbf{q}}+\text{h.c.}\right) \\
+\frac{V_{b}}{L^D} \sum_{\mathbf{kk^{\prime }q}}
a_{\mathbf{k+\frac{q}{2},} \uparrow }^{\dagger
}a_{\mathbf{-k+\frac{q}{2},}\downarrow }^{\dagger }a_{
\mathbf{-k^{\prime }+\frac{q}{2},}\downarrow }a_{\mathbf{k^{\prime
}+\frac{q }{2},}\uparrow }
\end{multline}%
where $\lambda _{b}$ (in units of $\hbar \omega$) is the relative
detuning , $\alpha _{b}$ (in units of $\hbar \omega a_{t}^{D/2}$) is
the coupling rate, and $V_{b}$ (in units of $\hbar \omega
a_{t}^{D}$) represents the remaining background interaction in the
open channel. We introduce the physical parameters related to three
bare parameters in $H_{\text{eff}}$ via a low-D renormalization
analogous to Eq. \eqref{eq:renormalization}:
\begin{gather}
V_{c}^{-1}=\int \frac{d^{D}\mathbf{k}}{\left( 2\pi \right) ^{D}}\frac{1}{%
2\epsilon _{\mathbf{k}}+3-D}\,,\quad \Omega ^{-1}=1-V_{p}V_{c}^{-1},  \notag
\\
V_{p}=\Omega ^{-1}V_{b}\,,\quad \alpha _{p}=\Omega ^{-1}\alpha
_{b}\,,\quad \lambda _{p}=\lambda _{b}-\Omega \alpha
_{p}^{2}V_{c}^{-1}\,. \label{eq:low-D renorm}
\end{gather}%
Note that the zero-point energy $3-D$ appears explicitly in the
definition of $V_{c}^{-1}$, otherwise there is an artificial
infrared divergence. These definitions of the physical parameters
are justified as we will see below that they remove exactly the
ultraviolet divergence associated with the contact interaction.

The effective Hamiltonian $H_{\text{eff}}$ should reproduce the same
physics represented by the 3D Hamiltonian $H$ when the system is
near the ground state with the energy per particle close to $E_{2B}
/2$ (as $\mu _{\rho }\ll 1$ in units of $\hbar \omega $). To
determine the parameters in $H_{\text{eff}}$, we match the exact
two-body bound state obtained from the original $H$. Specifically,
we first determine the effective background scattering $V_{p}$ by
matching the bound state physics in the deep BCS limit, where the
population is entirely atoms in the lowest trap mode. Then, for
general detuning, matching the binding energy and the
bound state gives two constraints which determine the remaining parameters, $%
\lambda _{b}$ and $\alpha _{b}$. Since the composition of the dressed
molecule $d_{\mathbf{q}}$ is a function of detuning $\nu _{p}$, or
equivalently, of the 3D scattering length, so are these two parameters.
\section{Fixing the parameters}
All the two-body physics contained in a given Hamiltonian are
embodied in its T-matrix, defined by $T\!\left( E \right) = H_{I} +
H_{I} G\!\left( E \right) H_{I} = H_{I} + H_{I} G_{0}\!\left( E
\right) T\!\left( E \right)$, where $H_{I}$ is the interaction part
of the Hamiltonian $H$, $G\!\left( E \right) = (E-H)^{-1}$ is the
full two-body propagator, and $G_{0}\!\left( E \right) =
(E-H_0)^{-1}$ is the free two-body propagator. Physically, the
matrix element $\langle \mathbf{2} | T \left( E \right) | \mathbf{1}
\rangle$ is the sum of the direct process whereby a pair of atoms
scatters from state $| \mathbf{1} \rangle$ to state $| \mathbf{2}
\rangle$ and the indirect processes whereby the atoms scatter into
an intermediate state, propagate with energy $E$, and then scatter
into the final state. From its definition, it is clear that the
T-matrix has simple poles where $E$ is equal to a two-body bound
state of the Hamiltonian. Furthermore, the residue of the diagonal
matrix element $\langle \mathbf{1} | T \left( E \right) | \mathbf{1}
\rangle$ at such a pole, $E=E_n$, is $|\langle \mathbf{1} |H_I|
\Psi_n \rangle|^2 = |\langle \mathbf{1} |H - H_0| \Psi_n \rangle|^2
= \left( E_n - \epsilon_{\mathbf{1}} \right)^2 |\langle \mathbf{1} |
\Psi_n \rangle|^2$, where $\epsilon_{\mathbf{1}}$ is the
noninteracting energy of state $| \mathbf{1} \rangle$. Thus, the
residue of a given diagonal matrix element determines the fraction
of the bound state in the given basis state. We shall use both of
these facts in matching the two-body bound state properties of the
exact and effective Hamiltonians.

In the appendix, we derive the general two-body T-matrix associated
with the Hamiltonian $H$ in Eq. \eqref{eq:H}. For a pair of atoms
asymptotically in the lowest mode of the trap, the corresponding
diagonal T-matrix element (in units of $a_{t}^{D}\hbar \omega $) as
a function of the two-body energy $E$ (measured with respect to the
continuum threshold, $3-D$) is
\begin{equation}
\left[ T\left( E\right) \right] ^{-1}=\gamma
_{\mathbf{000}}^{-2}\left( \left[ U_{p}^{\text{eff}}\left( E\right)
\right] ^{-1}-S_{p}\left( E\right) \right)\,, \label{eq:T}
\end{equation}%
where $\gamma_{\mathbf{000}}=\left( 2\pi \right) ^{\left( D-3\right)
/4}$, $U_{p}^{\text{eff}}\left( E\right)
\equiv U_{p}-g_{p}^{2}/(\nu _{p}-E)$ and, from the normalization in Eq. (2), $%
U_{p}^{\text{eff}}\left( E\right) =U_{b}^{\text{eff}}\left( E\right)
+U_{c}^{-1}\equiv U_{b}-g_{b}^{2}/\left( \nu _{b}-E\right) +U_{c}^{-1}$, and
the function
\begin{multline}
S_{p}\left( E\right) \equiv \frac{1}{L^D}\sum_{\mathbf{%
mnk}}\frac{\gamma _{\mathbf{mn0}}^{2}}{E-2\epsilon _{\mathbf{k}}-\varepsilon
_{\mathbf{m}}-\varepsilon _{\mathbf{n}}}+U_{c}^{-1} \\
=\frac{-1}{2^{5/2}\pi }%
\begin{cases}
\zeta \left( 1/2,-E/2\right) & \text{{\small {\emph{D}=1}}} \\
\int_{0}^{\infty }ds\left[ \frac{\Gamma \left( s-E/2\right) }{\Gamma \left(
s+1/2-E/2\right) }-\frac{1}{\sqrt{s}}\right] & \text{{\small {\emph{D}=2}}}%
\end{cases}
\label{eq:S}
\end{multline}%
\cite{sum note}.  In the above we have used the gamma function
$\Gamma \left( x\right) $ and the Hurwitz zeta function $\zeta
\left( s,x\right) =\lim_{N\rightarrow \infty }\sum_{n=0}^{N}\left(
n+x\right) ^{-s}-\left( N+x\right) ^{1-s}/(1-s)$.

The two-body bound state corresponds to a pole of the T-matrix
element above with $\left[ T\left( E_{2B}\right) \right] ^{-1}=0$ at
the binding energy $\left\vert E_{2B}\right\vert $, which gives the
eigen-equation
\begin{equation}
\left[ U_{p}^{\text{eff}}\left( E_{2B}\right) \right] ^{-1}=S_{p}\left(
E_{2B}\right)  \label{eq:energy}
\end{equation}
to determine $E_{2B}$.  Also, as discussed above, the atom pair
population of the lowest trap mode is determined by the residue of
the above T-matrix element at this pole. Since the bound state of
the \textit{effective} Hamiltonian comprises only atom pairs in the
lowest trap mode and dressed molecules, specifying the population in
the lowest mode is sufficient, in conjunction with normalization, to
determine the entire bound state of $H_{\text{eff}}$. Thus, to
ensure that the effective and the exact Hamiltonians produce the
same two-body bound state, we just need to obtain $\left[ T\left(
E\right) \right] ^{-1}$ from both $H$ and $H_{\text{eff}}$, and
require them to match to first order in $E-E_{2B}$.

Following the same approach as in the Appendix, the corresponding
T-matrix element from the effective Hamiltonian $H_{\text{eff}}$ is
obtained as
\begin{equation}
\left[ T^{\text{eff}}\left( E\right) \right] ^{-1}= \left[ V_{p}^{\text{eff}%
}\left( E\right) \right] ^{-1}- \gamma_{\mathbf{000}}^{-2} \sigma
_{p}\left( E\right), \label{eq:Teff}
\end{equation}%
where $\left[ V_{p}^{\text{eff}}\left( E\right) \right] ^{-1}\equiv
\left[ V_{p}-\alpha _{p}^{2}/\left( \lambda _{p}-E\right) \right] ^{-1}=%
\left[ V_{b}-\alpha _{b}^{2}/\left( \lambda _{b}-E\right) \right]
^{-1}+V_{c}^{-1}$, and the function%
\begin{align}
\sigma _{p}\left( E\right) &\equiv \gamma_{\mathbf{000}}^2 \int
\frac{d^{D}\mathbf{k}}{\left( 2\pi
\right) ^{D}}\left[ \frac{1}{E-2\epsilon _{\mathbf{k}}}+\frac{1}{2\epsilon _{%
\mathbf{k}}+3-D}\right] \notag
\\
&=\begin{cases}
\frac{-1}{4\pi\sqrt{-E}}+\frac{1}{2^{5/2} \pi} & D=1 \\
\frac{\ln \left( -E\right) }{2^{5/2} \pi^{3/2} } & D=2%
\end{cases}
\label{eq:sigma}
\end{align}

We require the effective background term alone to reproduce the
two-body physics on the deep BCS side, where the dressed molecule population is negligible.
Matching Eqs. \eqref{eq:T} and \eqref{eq:Teff} in that region yields
\begin{equation}
V_{p}^{-1}=\left( 2\pi \right) ^{\left( 3-D\right) /2}\left(
U_{p}^{-1}-C_{p}\right)  \label{eq:Vp}
\end{equation}%
where $C_{p}\equiv \lim_{\nu_p \rightarrow \infty} S_{p}\left(
E_{2B}\right) -\sigma _{p}\left( E_{2B}\right) $. In this way, we
obtain an effective single-channel model on the deep BCS side which
recovers the low-energy scattering models of Refs. \cite{3,4,7}.
However, the single-channel model with a renormalized scattering
length $\left( \propto V_{p}\right) $ is not adequate near resonance
or on the BEC side where the dressed molecule population becomes
significant.

To fix the parameters $\lambda _{p}$ and $\alpha _{p}$, we compare
the T-matrix in Eqs. \eqref{eq:T} and \eqref{eq:Teff} at general
detuning, and require them to match for the zeroth and the first
orders in expansion with $\left( E-E_{2B}\right) $ (which are
responsible for reproducing the same binding energy and the bound
state, respectively). After some algebra, we obtain
\begin{gather}
\lambda _{p}=E_{2B}\!-\!\frac{\sigma _{p}\left( E_{2B}\right) }{\partial
_{E_{2B}}\left[ U_{p}^{\text{eff}-1}-\left( S_{p}-\sigma _{p}\right) \right]
}\left[ 1-\frac{\sigma _{p}\left( E_{2B}\right) }{U_{p}^{-1}-C_{p}}\right]
\label{eq:lambda} \\
\alpha _{p}^{2}=\frac{\left( 2\pi \right) ^{\left( D-3\right)
/2}}{\partial _{E_{2B}}\left[ U_{p}^{\text{eff}-1}-\left(
S_{p}-\sigma _{p}\right) \right] }\left[ 1-\frac{\sigma _{p}\left(
E_{2B}\right) }{U_{p}^{-1}-C_{p}}\right] ^{2}  \label{eq:alpha}
\end{gather}
where $\partial _{E_{2B}}$ means $\partial /\partial E\big
|_{E=E_{2B}}$. Since the derivative of Eq. \eqref{eq:S} is not
obvious, we write it below explicitly \cite{dS note}.
\begin{equation}
\partial _{E_{2B}}S_{p}=\frac{-1}{2^{7/2}\pi }%
\begin{cases}
\frac{1}{2}\zeta \left( 3/2,-E_{2B}/2\right) & D=1 \\
\frac{\Gamma \left( -E_{2B}/2\right) }{\Gamma \left( 1/2-E_{2B}/2\right) } &
D=2%
\end{cases}
\label{eq:derivS}
\end{equation}

\begin{figure}[tbp]
\subfigure[$\lambda_p$]  {\includegraphics[width=120pt]
        {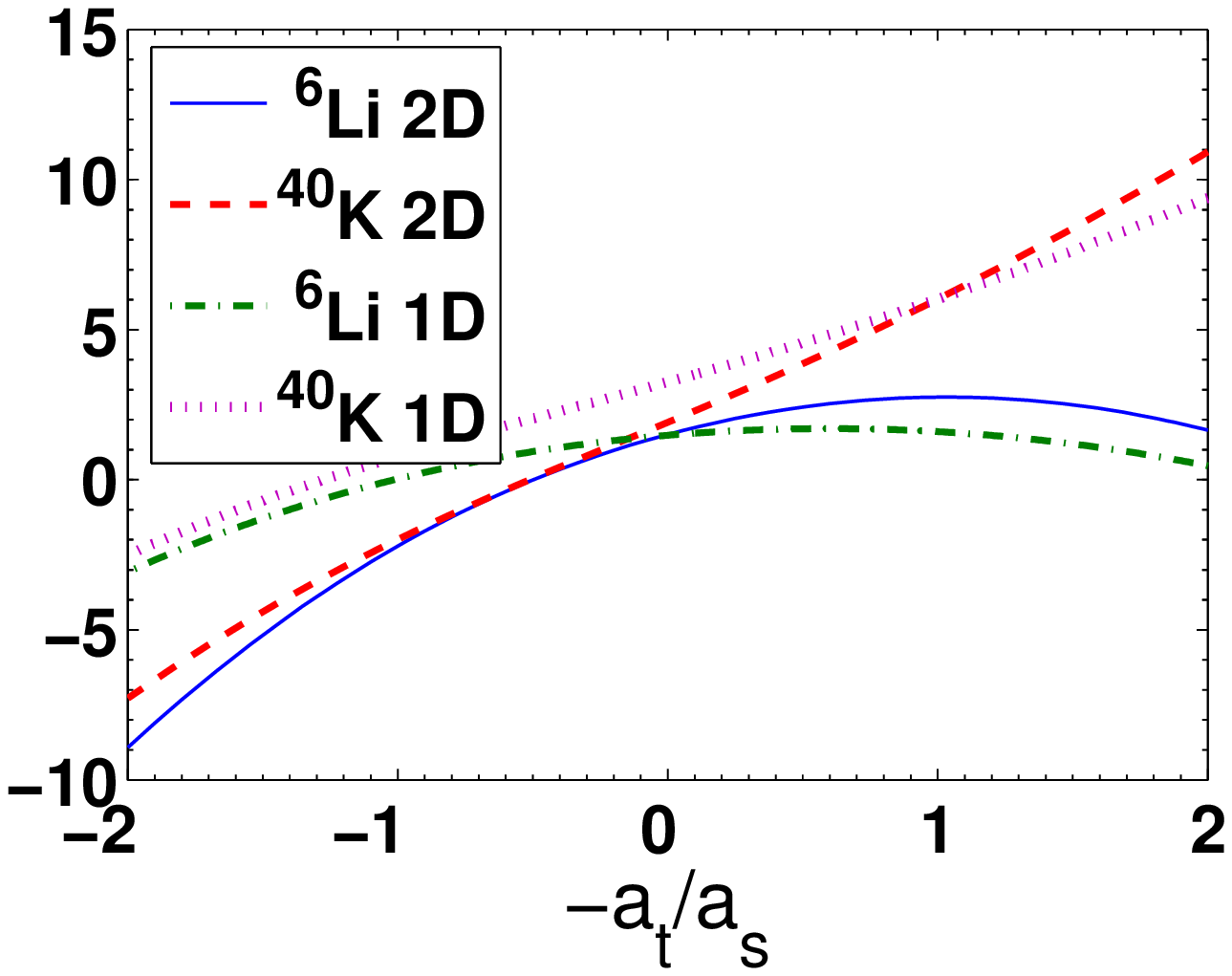}}
\subfigure[$\alpha_p^2$]  {\includegraphics[width=120pt]
        {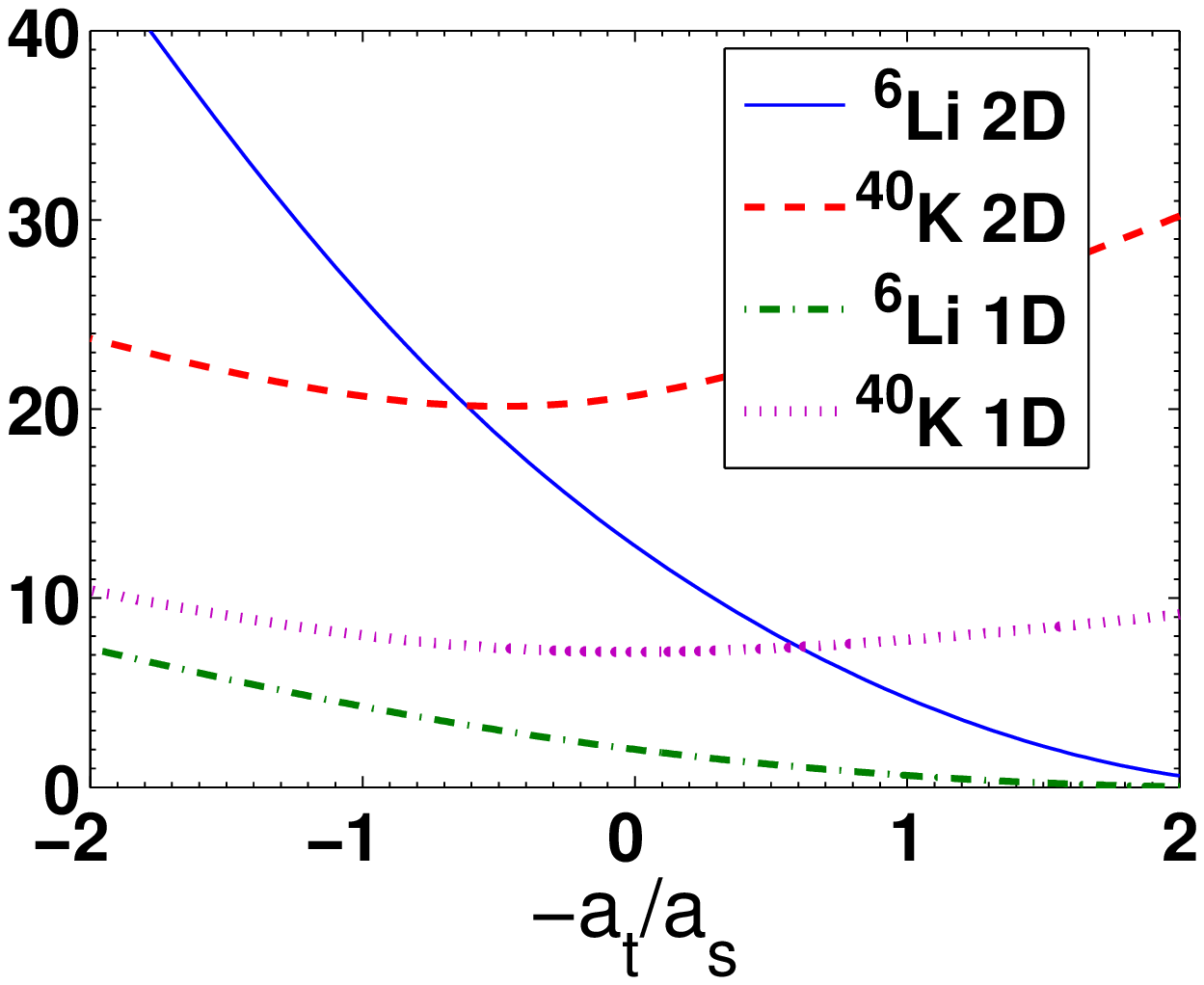}}
\subfigure[2D $V_p^{\text{eff}}$]  {\includegraphics[width=120pt]
        {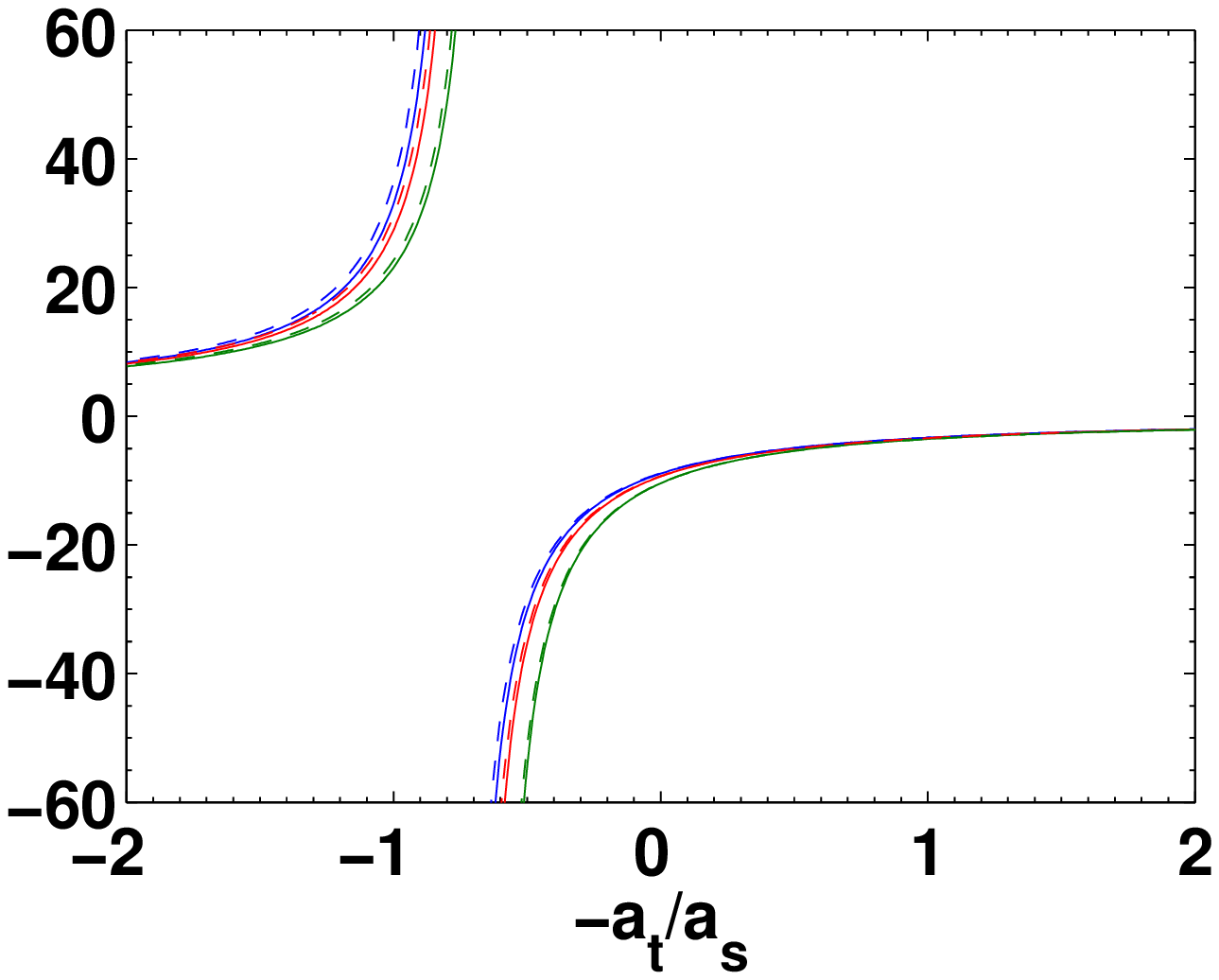}}
\subfigure[1D $V_p^{\text{eff}}$]  {\includegraphics[width=120pt]
        {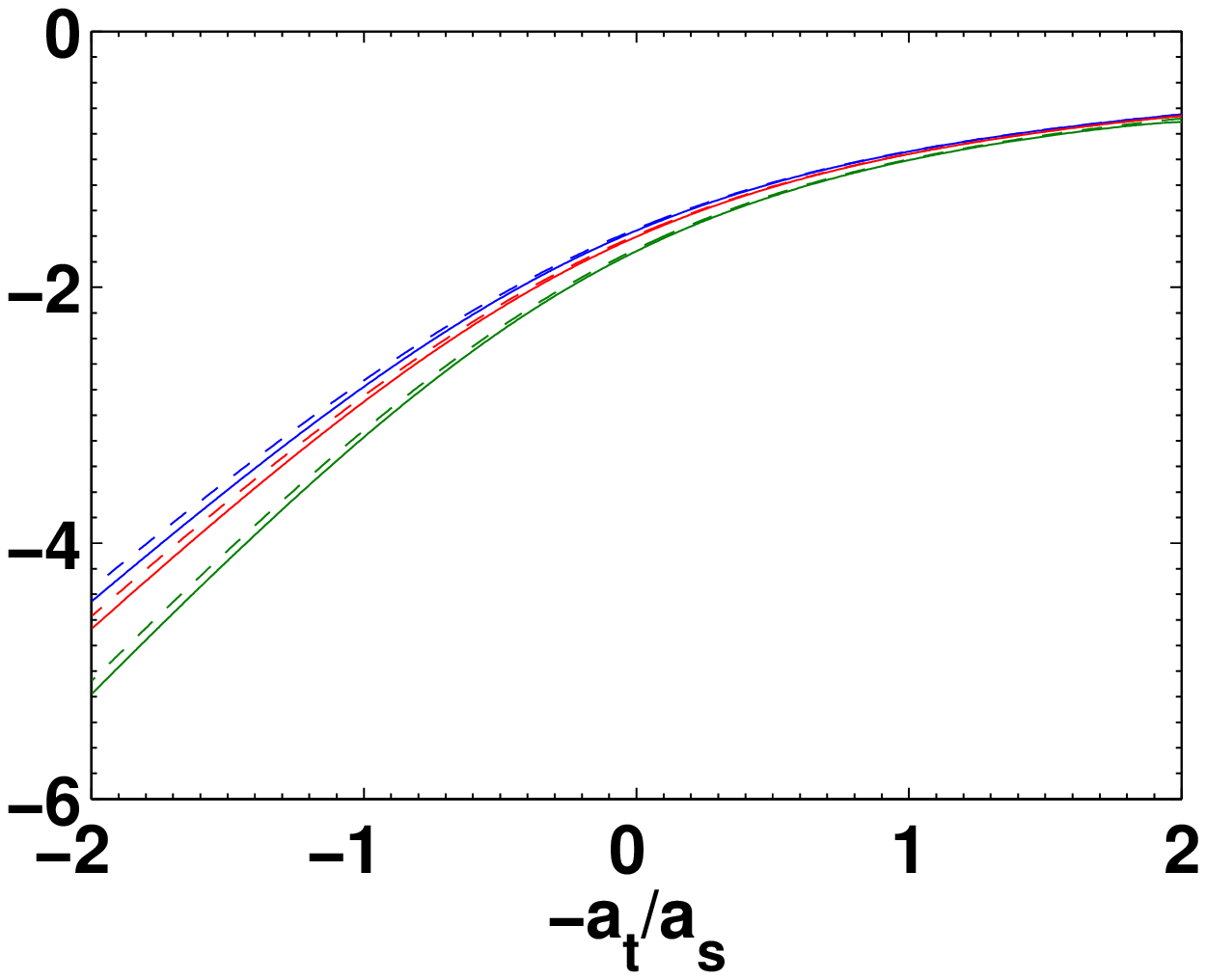}}
\caption{(Color online) (a)-(b): Detuning-dependent effective
parameters vs. inverse 3D scattering length, $a_t/a_s = 4\pi
/U_p^{\text{eff}} \left(0\right)$. (c)-(d): Effective interaction
$V_p^{\text{eff}} \left(2\mu\right)$ vs. $a_t/a_s$. Solid lines are
for $^{6}$Li and dashed lines are for $^{40}$K. The curves
correspond to $\protect\mu_{\rho} =0,0.05,0.15,$ from left to
right.}\label{fig:fig1}
\end{figure}

Thus, Eqs. (\ref{eq:Vp}, \ref{eq:lambda}, \ref{eq:alpha}) along with
the low-D renormalization procedure in Eq. \eqref{eq:low-D renorm}
fix the parameters of the effective Hamiltonian as functions of the
two-body binding energy, $E_{2B}$, which is connected to the
physical detuning $\nu_p$ through Eq. \eqref{eq:energy}. The
detuning-dependent parameters are plotted in Fig. \ref{fig:fig1}
across resonance, assuming the same typical 3D parameters as in Ref.
\cite{Kestner}. The difference in the sign of $U_{p}$ between
$^{6}$Li and $^{40}$K gives rise to quite different looking
effective parameters. However, the relevant combination of
parameters for mean-field calculations, $V_{p}^{\text{eff}}$, is
very nearly universal. We plot $V_{p}^{\text{eff}}\left( 2\mu
\right) $ across resonance with $\mu =E_{2B}/2+\mu _{\rho }$ for
small values of $\mu _{\rho }$ ($\mu $ has the meaning of the total
chemical potential including the per-particle binding energy
$E_{2B}/2$). The slight difference between the $^{6}$Li and $^{40}$K
curves stems from the slight difference in binding energies as a
function of the 3D scattering length. In the low density limit $\mu
_{\rho }\rightarrow 0$, $V_{p}^{\text{eff}}$ can approach infinity
in 2D on the BEC side where $\lambda _{p}=E_{2B}$, but in 1D
$V_{p}^{\text{eff}}$ is always attractive since we have $\lambda
_{p}>E_{2B}$ at any detuning. The 2D resonance-like behavior
($V_{p}^{\text{eff }}\left( E\right) \rightarrow \infty $) around
energy $E\sim E_{2B}$ should not be confused, however, with the
confinement induced resonance discussed in Ref. \cite{3} around
energy $E\sim 0$. Due to the existence of the two-body bound state
with $E_{2B}<0$, the resonance discussed in \cite{3} is not for the
ground-state configuration of the system.  Moreover, the relevant
quantity for a typical many-body calculation is $1/V_{p}^{
\text{eff}}\left( E\right)$, which is continuous.
\section{Summary}
In summary, we have derived an effective low-dimensional Hamiltonian
for strongly interacting atomic gas trapped in one or two dimensions
and free in the other dimensions. Excited trap modes are important
to the bound state physics, requiring the effective parameters to
assume a highly nontrivial magnetic field dependence. All the
parameters in the Hamiltonian have been fixed from two-body
considerations. This effective Hamiltonian can provide a starting
point to understand the low-dimensional many-body physics when the
system is near its ground state configuration (with the chemical
potential close to the per-particle binding energy $E_{2B}/2$).
\section{Acknowledgments}
This work was supported by the MURI, the DARPA, the NSF award
(0431476), the DTO under ARO contracts, and the A. P. Sloan
Foundation. J.P.K. gratefully acknowledges many helpful discussions
with Wei Zhang and G.-D. Lin.

\section{Appendix}
Below we derive the center-of-mass two-body T-matrix between atoms
in arbitrary trap modes from the 3D Hamiltonian $H$ in Eq. (1). We
use $|\mathbf{mnk}\rangle $ to denote an atomic pair state in trap
modes $\mathbf{m}$ and $\mathbf{n}$ having relative momentum
$\mathbf{k}$, and $|\mathbf{p}\rangle $ to denote a molecule in trap
mode $\mathbf{p}$. With the definition $T=H_{I}+H_{I}G_{0}T$, we
evaluate the propagator at energy $E$ and obtain
\begin{multline}
\langle \mathbf{mnk}|T|\mathbf{m^{\prime }n^{\prime }k^{\prime }}\rangle
=U_{b}\gamma _{\mathbf{mn}}^{\mathbf{m^{\prime }n^{\prime }}}\!+\sum_{%
\mathbf{p}}\frac{g_{b}\gamma _{\mathbf{mnp}}\langle \mathbf{p}|T|\mathbf{%
m^{\prime }n^{\prime }k^{\prime }}\rangle }{E-\varepsilon _{\mathbf{p}}-\nu
_{b}}  \label{eq:L-S open1} \\
+\sum_{\mathbf{m^{\prime \prime }n^{\prime \prime }k^{\prime \prime }}}\frac{%
U_{b}\gamma _{\mathbf{mn}}^{\mathbf{m^{\prime \prime }n^{\prime \prime }}%
}\langle \mathbf{m^{\prime \prime }n^{\prime \prime }k^{\prime \prime }}|T|%
\mathbf{m^{\prime }n^{\prime }k^{\prime }}\rangle }{E-2\epsilon _{\mathbf{%
k^{\prime \prime }}}-\varepsilon _{\mathbf{m^{\prime \prime }}}-\varepsilon
_{\mathbf{n^{\prime \prime }}}}
\end{multline}
\begin{equation}
\langle \mathbf{p}|T|\mathbf{mnk}\rangle =g_{b}\gamma _{\mathbf{mnp}%
}+\!\!\!\!\!\sum_{\mathbf{m^{\prime \prime }n^{\prime \prime }k^{\prime
\prime }}}\!\!\!\!\frac{g_{b}\gamma _{\mathbf{m^{\prime \prime }n^{\prime
\prime }p}}\langle \mathbf{m^{\prime \prime }n^{\prime \prime }k^{\prime
\prime }}|T|\mathbf{mnk}\rangle }{E-2\epsilon _{\mathbf{k^{\prime \prime }}%
}-\varepsilon _{\mathbf{m^{\prime \prime }}}-\varepsilon _{\mathbf{n^{\prime
\prime }}}}  \label{eq:L-S open-closed}
\end{equation}
Substituting Eq. \eqref{eq:L-S open-closed} into Eq. \eqref{eq:L-S
open1}, we get
\begin{multline}
\langle \mathbf{mnk}|T|\mathbf{m^{\prime }n^{\prime }k^{\prime }}\rangle
=\sum_{\mathbf{p}}U_{b}^{\text{eff}}\left( E-\varepsilon _{\mathbf{p}%
}\right) \gamma _{\mathbf{mnp}}\biggl[\gamma _{\mathbf{m^{\prime }n^{\prime
}p}}  \label{eq:L-S open2} \\
+\sum_{\mathbf{m^{\prime \prime }n^{\prime \prime }k^{\prime \prime }}}\frac{%
\gamma _{\mathbf{m^{\prime \prime }n^{\prime \prime }p}}}{E-2\epsilon _{%
\mathbf{k^{\prime \prime }}}-\varepsilon _{\mathbf{m^{\prime \prime }}%
}-\varepsilon _{\mathbf{n^{\prime \prime }}}}\langle \mathbf{m^{\prime
\prime }n^{\prime \prime }k^{\prime \prime }}|T|\mathbf{m^{\prime }n^{\prime
}k^{\prime }}\rangle \biggr]
\end{multline}
where $U_{b}^{\text{eff}}\left( E\right) $ is defined in the text.
Note that we can separate out the final state dependence
\begin{equation}
\langle \mathbf{mnk}|T|\mathbf{m^{\prime }n^{\prime }k^{\prime }}\rangle
=\sum_{\mathbf{p}}U_{b}^{\text{eff}}\left( E-\varepsilon _{\mathbf{p}%
}\right) \gamma _{\mathbf{mnp}}\tilde{T}_{\mathbf{m^{\prime }n^{\prime
}k^{\prime }}}^{\mathbf{p}}
\end{equation}
Substituting this form into Eq. \eqref{eq:L-S open2}, we get
\begin{multline}
\tilde{T}_{\mathbf{m^{\prime }n^{\prime }k^{\prime }}}^{\mathbf{p}}=\gamma _{%
\mathbf{m^{\prime }n^{\prime }p}}+\sum_{\mathbf{p^{\prime }}}U_{b}^{\text{eff%
}}\left( E-\varepsilon _{\mathbf{p^{\prime }}}\right)
\tilde{T}_{\mathbf{m^{\prime
}n^{\prime }k^{\prime }}}^{\mathbf{p^{\prime }}}  \label{eq:L-S open3} \\
\times \sum_{\mathbf{m^{\prime \prime }n^{\prime \prime }k^{\prime \prime }}}%
\frac{\gamma _{\mathbf{m^{\prime \prime }n^{\prime \prime }p}}\gamma _{%
\mathbf{m^{\prime \prime }n^{\prime \prime }p^{\prime }}}}{E-2\epsilon _{%
\mathbf{k^{\prime \prime }}}-\varepsilon _{\mathbf{m^{\prime \prime }}%
}-\varepsilon _{\mathbf{n^{\prime \prime }}}}
\end{multline}
The inner sum is equal to $S\left( E-\varepsilon
_{\mathbf{p}}\right) \delta _{\mathbf{p^{\prime }}\mathbf{p}}$
\cite{sum note}, where $S\left( E\right) \equiv S_{p}\left( E\right)
-U_{c}^{-1}$ and $S_{p}\left( E\right) $ is defined as in Eq.
\eqref{eq:S}, so we can solve Eq. \eqref{eq:L-S open3} to obtain
\begin{equation}
\tilde{T}_{\mathbf{m^{\prime }n^{\prime }k^{\prime }}}^{\mathbf{p}%
}=\frac{\gamma _{\mathbf{m^{\prime }n^{\prime }p}}}{1-U_{b}^{\text{eff}%
}\left( E-\varepsilon _{\mathbf{p}}\right) S\left( E-\varepsilon _{\mathbf{p}%
}\right) }
\end{equation}
and
\begin{equation}
\langle \mathbf{mnk}|T|\mathbf{m^{\prime }n^{\prime }k^{\prime }}\rangle
=\sum_{\mathbf{p}}\frac{\gamma _{\mathbf{mnp}}\gamma _{\mathbf{m^{\prime
}n^{\prime }p}}}{\left[ U_{p}^{\text{eff}}\left( E-\varepsilon _{\mathbf{p}%
}\right) \right] ^{-1}-S_{p}\left( E-\varepsilon _{\mathbf{p}}\right) }
\label{eq:T open-open}
\end{equation}%
\begin{equation}
\langle \mathbf{p}|T|\mathbf{mnk}\rangle =\frac{g_{b}\gamma _{\mathbf{mnp}}}{%
1-U_{b}^{\text{eff}}\left( E-\varepsilon _{\mathbf{p}}\right) S\left(
E-\varepsilon _{\mathbf{p}}\right) }  \label{eq:T open-closed}
\end{equation}

For the special case of atoms asymptotically in the lowest mode of
the trap, $\mathbf{m} = \mathbf{n} = \mathbf{m^{\prime }} =
\mathbf{n^{\prime }} = \mathbf{0}$, we
use the fact that $\gamma_{\mathbf{00p}} \propto \delta_{\mathbf{p}, \mathbf{%
0}}$ to obtain Eq. \eqref{eq:T} from Eq. \eqref{eq:T open-open}.


\begin{thebibliography}{99}
\bibitem{dS note} We first differentiate the summand in Eq. \eqref{eq:S}, then sum.
\bibitem{sum note} Rewriting the numerator with the explicit
definitions \eqref{eq:gamma} and \eqref{eq:phi} and using the
identity $\frac{1}{x} = -\int_0^{\infty} dt e^{t x}$, valid for
$x<0$, to rewrite the denominator, the sum can be written as a [$2
\times \left( 3-D \right) + 1$]-dimensional integral of a more
convenient sum which can be done analytically using Mehler's Hermite
polynomial formula. The spatial integrals can then be performed,
giving rise to the Kronecker delta. Finally, after integrating over
the dummy variable $t$, the results can be analytically continued to
$x>0$.
\end{thebibliography}
\end{document}